\documentclass[12pt,preprint]{aastex}

\newcommand{\agn}{{\small AGN}}
\newcommand{\ngc}{{\small NGC~3783}}
\newcommand{\uta}{{\small UTA}}
\newcommand{\hullac}{{\small HULLAC}}

\newcommand{\xmm}{{\it XMM-Newton}}
\newcommand{\rgs}{{\small RGS}}
\newcommand{\chandra}{{\it Chandra}}

\newcommand{\hetgs}{{\small HETGS}}

\newcommand{\uv}{{\small UV}}
\newcommand{\x}{X-ray}
\newcommand{\xs}{X-rays}
\newcommand{\kms}{km~s$^{-1}$}
\begin{document}

\title{Is The Fe M-shell Absorber Part of The Outflow in Active Galactic Nuclei?}
\author{Tomer Holczer \altaffilmark{1}, Ehud Behar \altaffilmark{1} and
Shai Kaspi \altaffilmark{1}}

\altaffiltext{1}{Department of Physics,
                 Technion, Haifa 32000, Israel.
                 tomer@physics.technion.ac.il (TH),
                 behar@physics.technion.ac.il (EB),
                 shai@physics.technion.ac.il (SK).}
\received{} \revised{7/1/04} \accepted{}

\shorttitle{Is The Fe M-shell Absorber Part of The Outflow?}
\shortauthors{Holczer et al.}

\begin{abstract}

  The \x\ emission of many active galactic nuclei (\agn s)
  is absorbed between 15~and 17~\AA\ by means of
  unresolved (inner-shell) transition arrays (\uta s) of Fe M-shell ions.
  The outflow velocities implied by the Doppler shifts of the
  individual \uta s in the spectrum have never before been measured.
  Thus,  the Fe-M absorber has been commonly
  assumed to be part of the ionized \agn\ outflow, whose
  velocities are readily obtained from more easily measured spectral lines.
  The best spectrum of Fe-M absorption is available from the integrated 900~ks \chandra\ \hetgs\
  observations of \ngc, in which some Fe-M ions are clearly resolved.
  We measure the velocities of the individual Fe-M ions in \ngc\ for the first time.
  Surprisingly, we find that the Fe-M absorber, most noticeably Fe$^{+8}$, Fe$^{+9}$, and
  Fe$^{+10}$, is not outflowing at the same velocity as the previously known wind.
  In fact, it appears to be stationary and therefore not part of the outflow at all.
  It could, alternatively, be ascribed to the skin of the dusty torus.
  This reduces appreciably the mass loss rate estimated for the \ngc\ outflow
  and perhaps for other similar sources as well, in which the various Fe-M ions are not
  resolved.

\end{abstract}


\section{INTRODUCTION}
\label{sec:intro}

The \x\ spectrum of many active galactic nuclei (\agn s) viewed
directly towards the central source (e.g., Seyfert~1 galaxies)
shows the continuum flux absorbed by numerous absorption lines.
These lines are generally shifted slightly to shorter wavelengths
indicating an outflowing wind. The flux from the central source is
believed to drive the wind and to ionize it to the high charge
states observed. The \x\ band is uniquely compact as it comprises
absorption lines from the full range of charge states from neutral
atoms to H-like ions of all elements with $Z~\geq$ 6, which helps
to constrain the column density distribution over a wide range in
ionization. The less ionized species absorb by means of
inner-shell photoexcitation processes, which play an important
role in the formation of discrete features in the spectrum.

In particular, the $n$~= 2 to 3 absorption lines ($n$ being the
principal quantum number of the active electron) by low charge
states of Fe, neutral through Fe$^{+15}$, fall in the soft X-ray
band. Each ion can produce dozens of overlapping lines forming
unresolved transition arrays (\uta s). \citet{sako01} reported the
first astrophysical identification of these features in the
spectrum of the luminous quasar IRAS 13349+2438 obtained with the
Reflection Grating Spectrometer (RGS) on board \xmm. The broad
absorption trough observed between 16--17~\AA\ was attributed to a
blend of lines from Fe$^{+6}$ to Fe$^{+11}$.
The basic atomic data needed for spectral modeling of the lines
was subsequently given in \citet{behar01}. Since its initial
discovery, copious detections of Fe M-shell absorption have been
claimed in other soft \x\ sources \citep[e.g.,][to name a few
recent ones]{pounds04,kaspi04,gallo04}. Since most instruments are
not able to resolve the various ionic \uta s, for simplicity, the
observed, blended feature was dubbed in the literature as the Fe-M
\uta.

Perhaps the most conspicuous astrophysical Fe-M \uta\ is the one
found in \ngc. Both \chandra\ and \xmm\ have observed \ngc\ for
very long spectroscopic exposures
\citep[][respectively]{kaspi02,behar03}. Several detailed global
models have been constructed to explain the rich absorption
spectrum of \ngc\ \citep{kaspi01,blustin02,krongold03,netzer03}.
Owing to its ionization sensitivity, the Fe-M \uta\ could have
played a major role in constraining these models. However, it was
realized that the standard photoionization balance models fail
tremendously in their prediction of the Fe charge state
distribution \citep{netzer03,netzer04,kraemer04}. Therefore, the
current models could not produce the Fe-M \uta\ spectral features
consistently with those of other elements. Recently,
\citet{krongold05} have used heavy rebinning of the \chandra\
spectrum of \ngc\ to show that the overall position of the
absorption trough in the Fe-M \uta\ region appears to shift
between low and high states of the ionizing continuum. In none of
these works were the individual ionic features studied closely.

In this paper, we revisit the integrated 900~ks observation of
\ngc\ obtained with the \hetgs\ spectrometer on board \chandra\
over several months. The same data were used before by
\citet{kaspi02}, \citet{netzer03}, and by \citet{krongold03}.
Presently, \hetgs\ is the only instrument that can actually
resolve the individual ionic features inside the Fe-M \uta.
Conversely, the spectral resolution of the \rgs\ on board \xmm\ is
just insufficient \citep{sako01,behar03}. Curiously, the details
of the \uta\ in the \hetgs\ spectrum have yet to be studied in
detail, perhaps because of the difficulties mentioned above in
using the Fe-M \uta\ in global models. The sole goal of this work
is to use the \hetgs\ spectrum to identify the individual ionic
features and to measure their velocity with respect to the
systemic velocity of \ngc. The results obtained are quite
surprising.


\section{HETGS SPECTRUM OF NGC 3783}
\label{sec:data}

The spectrum used in the present work is the exact same one
reduced in \citet{kaspi02} and shown in their Fig.~1. It consists
of five 170~ks \hetgs\ observations of \ngc\ obtained during the
period 2001 February--June as well as an early 56~ks observation
from 2000 January, all of which are co-added. The +1st and $-$1st
diffraction orders of both the MEG and HEG grating assemblies are
included. The spectra are fluxed and corrected for Galactic
absorption. The wavelength scale is de-redshifted to the rest
frame of \ngc\ using z~= 0.00976. For more details on the data
reduction see \citet{kaspi02}. For the obvious reason of improving
the statistics, we use here the total 900~ks data set.
Notwithstanding, we have verified that all of the prominent
absorption features, which we discuss below, appear in each short
exposure. Moreover, to the extent that it is possible to determine
with the low statistics of each short exposure, we have found that
the positions of the narrow absorption features, which are of
interest for this work, do not vary between observations.

\section{SPECTRAL MODEL}
\label{sec:modeling}

The heavy line blending in the soft \x\ band where the outflow
absorbs most significantly makes the direct measurement of
individual line positions very difficult. For a consistent
measurement, therefore, one needs to use an ad-hoc spectral model,
which includes simultaneously many charge states and all of the
absorption lines pertaining to these ions in the soft \x\ band.
Interested solely in the absorption features of the Fe-M \uta, we
have constructed a simple model to explain the observed spectrum
in the limited range of 14.9--17.5~\AA. The model includes a
power-law continuum, all of the Fe-M \uta\ ions, Fe$^{+16}$, and
the two O-K ions: O$^{+6}$ and O$^{+7}$, which also have lines in
this range. For the Fe ions, the model includes all of the $n$~= 2
to 3 spectral lines for absorption from the ground level of each
ion. The present Fe-M model includes the full set of absorption
lines from \citet{behar01} and not only the abbreviated set given
in Table~1 of that paper. For the O-K ions, the model includes all
of the $n$~= 1 to $n'$ resonance transitions with $n'~\leq$ 15000
as well as the photoelectric edges, from the ground level. The
present model is similar in spirit to those used in \citet{sako01}
and in \citet{behar03}.

For the continuum, a power law was fit to the relatively
featureless region of the spectrum between 2--6~\AA. The best-fit
parameters obtained are a photon index ($\Gamma$) of 1.45 and a
normalization at 1~keV of 0.00936
photons~s$^{-1}$~cm$^{-2}$~keV$^{-1}$, which are in good agreement
with previous fits for the low-state spectrum that dominates the
present data set \citep[e.g., Fig.~8 in][]{netzer03}.

The optical depth $\tau_{ij}(\nu)$ due to an absorption line
($i~\to j$) is obtained by:

\begin{equation}
\tau_{ij}(\nu) = N_{ion}\sigma_{ij}(\nu)
\end{equation}

\noindent where $\sigma_{ij}(\nu)$ is the calculated cross section
for photoexcitation (in cm$^2$) from $i$ to $j$ and $N_{ion}$ is
the ionic column density towards the source (in cm$^{-2}$) for
which we fit. The photoexcitation cross section is given by:

\begin{equation}
\sigma_{ij}(\nu) = \frac{\pi e^2}{m_ec}f_{ij}\phi(\nu)
\end{equation}

\noindent Here, $e$ and $m_e$ are the electron charge and mass,
$c$ is the speed of light, $f_{ij}$ is the line oscillator
strength, which we have calculated with the Hebrew University
Lawrence Livermore Atomic Code \citep[\hullac,][]{bs01}. Finally,
$\phi(\nu)$ is the normalized line profile described here by a
Voigt function that is the convolution of the (Gaussian) Doppler
broadening due to temperature and to turbulence, which we fit for,
and the (Lorentzian) natural width of the line, which we have
calculated with \hullac. In observed \agn\ outflows, turbulence
broadening ($v_{turb}~\ge$~100~\kms) totally dominates over
temperature broadening. Here, we use $v_{turb}$~= 170~\kms.
Generally, in the natural widths of inner-shell excited lines, the
autoionization decay rate ($A^a\simeq$ 10$^{14}$~s$^{-1}$)
dominates over the radiative rate. Nevertheless, our calculations
include both. \hullac\ is a multi-configuration, relativistic code
most suitable for the inner-shell excitations of the complex atoms
of interest in this study. For more details on the Fe-M atomic
calculations see \citet{behar01}.

\section{DETERMINATION OF OUTFLOW VELOCITIES}
\label{sec:velocity}

The mean velocity for the \ngc\ outflow was found by
\citet{kaspi02} to be $-$590 $\pm$~150~\kms. Consequently, our
first attempt to fit the absorption in the \hetgs\ spectrum of
\ngc\ in the vicinity of the Fe-M \uta\ was carried out using a
uniform outflow velocity of --590~\kms. The results are shown in
Fig.~1. Indeed, the fit for O$^{+6}$ (He~$\delta$ and
He~$\epsilon$ most prominent at 17.40 and 17.20~\AA), O$^{+7}$
(Ly~$\beta$ and Ly~$\gamma$ at 16.01 and 15.18~\AA), and
Fe$^{+16}$ (at 15.01 and 15.26~\AA) is seen to be very good. On
the other hand, the velocity of --590~\kms\ is clearly
inappropriate for the prominent Fe-M troughs. The Fe-M features in
the model are seen to be systematically shifted towards shorter
wavelengths compared with their observed positions, most evidently
at 16.50, 16.33, and 16.15~\AA.

The above result is already a strong indication that the Fe-M
absorber is moving at a much slower velocity than the more highly
ionized gas, if it is moving at all. In Table~1 we list the
identifiable discrete absorption features in the spectrum. For the
mere objective of obtaining centroids in order to get a rough idea
of the velocities, we have fitted each feature with a local
continuum and a Gaussian-shaped trough employing the same method
used for the wavelength determination in \citet{kaspi02}. It can
be seen that the positions of the best observed Fe-M ions are
consistent with outflow velocities that are significantly lower
than $-$~590~\kms. The velocities in the table have to be treated
with caution for two main reasons: (i) The absorption line profile
in many cases is not a simple Gaussian, but rather a complex
distribution of narrow velocity components \citep[see
also][]{kaspi02, gabel03} (ii) The Fe-M wavelengths have not been
confirmed in the laboratory as discussed in detail in \S5. The
spectral resolution of the \hetgs\ is high enough so that blending
among Fe-M ions is negligible as can be seen by comparing the 3rd
and 4th columns in Table~1, which show that the individual-ion
Fe-M wavelengths and the full-model wavelengths agree to within a
few m\AA.

\begin{deluxetable}{lccccc}
\tablecolumns{5} \tablewidth{0pt}
\tablecaption{Best-fit velocities and column densities for ions
detected in the 14.9--17.5~\AA\ region of the \hetgs\ spectrum of
\ngc. \label{tab1}} \tablehead{
   \colhead{Ion} &
   \colhead{$\lambda _{\mathrm {Observed}}$} &
   \colhead{$\lambda _{\mathrm {Rest}}$\tablenotemark{a}} &
   \colhead{$\lambda _{\mathrm {Model}}$\tablenotemark{b}} &
   \colhead{Outflow Velocity \tablenotemark{c}} &
   \colhead{Ion Column Density } \\
   \colhead{} &
   \colhead{(\AA)} &
   \colhead{(\AA)} &
   \colhead{(\AA)} &
   \colhead{(\kms)} &
   \colhead{(10$^{16}$~cm$^{-2}$)}
}
  \startdata
O$^{+7}$ & 15.144 $\pm$ 0.003 \tablenotemark{d} & 15.176 & 15.146 & $-$632 $\pm$ 59 & 400 $\pm$ 60 \\
         & 15.970 $\pm$ 0.005 \tablenotemark{e} & 16.006 & 15.973 & $-$665 $\pm$ 94 & \\
O$^{+6}$ & 17.351 $\pm$ 0.005 & 17.395 & 17.361 & $-$770 $\pm$ 81 & 110 $\pm$ 20 \\
         & 17.161 $\pm$ 0.005 & 17.199 & 17.165 & $-$675 $\pm$ 89 &   \\
         & 17.048 $\pm$ 0.005 & 17.084 & 17.050 & $-$630$ \pm$ 88  & \\
Fe$^{+16}$ & 14.980 $\pm$~0.003 & 15.013 & 14.985 & $-$659 $\pm$ 60 & 3.0 $\pm$ 0.5  \\
           & 15.231 $\pm$~0.002 \tablenotemark{f} & 15.261 & 15.234 & $-$590 $\pm$ 39 & \\
\cline{1-6}
Fe$^{+15}$ & 15.231 $\pm$~0.002 \tablenotemark{f} & 15.250 & 15.234 & $-$374  $\pm$ 39 & 0.6 $\pm$ 0.2 \\
Fe$^{+14}$ & 15.322 $\pm$ 0.006 & 15.316 & 15.317 & $+$118 $\pm$ 118 & 0.3 $\pm$ 0.1 \\
Fe$^{+13} $ & 15.569 $\pm$ 0.008 \tablenotemark{g} & 15.580 & 15.580 & $+$212 $\pm$ 154 & 1.4 $\pm$ 0.4  \\
Fe$^{+12} $ & 15.844 $\pm$ 0.014 \tablenotemark{g} & 15.848 & 15.848 & $-$76 $\pm$ 265 & 1.0 $\pm$ 0.3  \\
Fe$^{+11}$ & 15.970 $\pm$ 0.005 \tablenotemark{e} & 15.967 & 15.973 & $+$56 $\pm$ 94 & 3.0 $\pm$ 2.0 \\
Fe$^{+10}$ & 16.154 $\pm$ 0.005 & 16.150 & 16.150 & $+$74 $\pm$ 93 & 4.0 $\pm$ 0.7 \\
Fe$^{+9}$ & 16.329 $\pm$ 0.005 & 16.339 & 16.340 & $-$184 $\pm$ 92 & 5.5 $\pm$ 0.7  \\
Fe$^{+8}$ & 16.496 $\pm$ 0.004 & 16.510 & 16.509 & $-$252 $\pm$ 73 & 4.0 $\pm$ 0.5 \\
Fe$^{+7}$ \tablenotemark{h} & \nodata & 16.655 & \nodata & \nodata & 3.0 $\pm$ 1.0 \\
Fe$^{+6}$ \tablenotemark{h} & \nodata & 17.095 & \nodata & \nodata & 2.0 $\pm$ 0.7 \\
Fe$^{+5}$ \tablenotemark{h} & \nodata & 17.218 & \nodata & \nodata & 1.5 $\pm$ 0.6 \\
Fe$^{+4}$ \tablenotemark{h} & \nodata & 17.291 & \nodata & \nodata & 1.0 $\pm$ 0.6 \\
Fe$^{+3}$ \tablenotemark{h} & \nodata & 17.405 & \nodata & \nodata & $\leq$ 0.8  \\
%
\enddata
\footnotesize \tablenotetext{a}{ For Fe-M ions, these are
centroids of the deepest feature in each ionic spectrum.}

\tablenotetext{b} { Centroid in full two-velocity model (Fig.~2).}

\tablenotetext{c} { Estimated by ($\lambda _{\mathrm Observed} -
\lambda _{\mathrm Rest}) c/ \lambda _{\mathrm Rest}$. Errors
reflect 90\% confidence intervals.}

\tablenotetext{d} { Blend of O$^{+7}$ and Fe$^{+15}$ at 15.19~\AA\
fitted for two distinct features.}

\tablenotetext{e}{ Unresolved blend of O$^{+7}$ and Fe$^{+11}$
reflected in the large error on the Fe$^{+11}$ column density.}

\tablenotetext{f}{ Unresolved blend of Fe$^{+16}$ and Fe$^{+15}$.}

\tablenotetext{g}{ Large uncertainty in centroid identification.
See spectrum in Figs.~1 and 2. }

\tablenotetext{h}{ Indirect identification based on best-fit
model; Strongly blended with the high-$n$ lines of O$^{+6}$.}
\end{deluxetable}

In order to further illustrate the low velocities associated with
the Fe-M absorber, we have fitted the \hetgs\ spectrum of \ngc,
but now the velocity of the Fe-M ions was allowed to be different
from that of the other charge states. For simplicity, however, the
velocities of all of the Fe-M ions were tied together. The best
fit was obtained by setting the Fe-M velocity to zero as shown in
Fig.~2. The individual contributions of Fe$^{+7}$ to Fe$^{+12}$ to
the total model are plotted in Fig.~3. Despite some blending,
individual ions can be easily discerned in the spectrum. As can be
seen, centroids of the strongest absorption features in the model
are determined almost exclusively by a single \uta, which
reinforces the validity of the direct velocity estimates in
Table~1. The most conspicuous Fe-M ions in the data are Fe$^{+8}$,
Fe$^{+9}$, and Fe$^{+10}$. These ions in particular produce \uta s
that are only very weakly blended with O lines or with other Fe
ions as demonstrated in Fig.~3. Particularly for these ions, the
fit of the model with zero outflow velocity can be seen in Fig.~2
to be very good. The obvious velocity discrepancies for these ions
in Fig.~1 have now disappeared. Conclusions regarding higher and
lower Fe-M charge states are not as obvious, although the data
still favor the stationary scenario over the $-$~590~\kms\ outflow
for these ions as well, as was already indicated by the direct
velocity estimates presented in Table~1.

\section{DISCUSSION}
\label{sec:discus}

The outflow of \ngc\ is known from high-resolution \uv\
observations to have four resolved velocity components
\citep{gabel03}. In the \xs, the four components are unresolved.
However, the mean \x\ absorption line profile is generally in
agreement with the \uv\ profile, as demonstrated in Fig.~2 of
\citet{gabel03}. The lowest velocity \uv\ absorber reported in
\citet{gabel03} is at $-$548~\kms, where the absorption troughs
rise sharply. On the other hand, from our own inspection of the
HST and FUSE spectra of \ngc\ we find a much slower absorber, the
most prominent lines of which are Ly$\alpha$ and Ly$\beta$
observed at an infall velocity of about $+$100~\kms\ \citep[see
also Fig.~1 in][]{gabel03}.  As far as we know, this together with
the present \x\ results represents the first discovery of a high
column density, almost stationary \x\ component in any \agn. In
the intensive \xmm\ study of \ngc, the low charge states of oxygen
(neutral O and O$^{+1}$) have actually been detected at zero
velocity as well \citep[][Table 1 therein]{behar03}. However, both
lines were stated to be blends and in any event neither feature
was statistically significant. The present detection is much more
unambiguous.

The stationary or low-velocity component could possibly be an
extended, slightly ionized part of the interstellar medium (ISM)
of \ngc. It might be absorption created by the extended narrow
line region of \ngc. It could also be gas evaporating off of the
molecular torus. A scenario in which radiation pressure is
inefficient has been suggested for \ngc\ by \citet{chelouche05}.
The outflow velocity has obvious consequences on the mass outflow
rate, which for \ngc\ has been estimated to exceed a solar mass
per year \citep{behar03}. In fact, \citet{netzer03} have estimated
the mass loss rate in this target due to the low ionization
component alone to be as high as 75 solar masses per year and
scaling linearly with the radial filling factor of the flow, which
could be as low as 0.1 \citep{netzer03}. All of these estimates
for the mass loss rate are based on the high velocity of the
outflow ($\sim~-$590~\kms). If indeed, much of this gas is not
moving as fast, it would reduce the estimated mass loss rate
considerably.

Outflow velocities for individual Fe-M ions have not been carried
out for other \agn\ targets, mostly because the available
observations generally do not resolve the individual ions in the
spectrum, even when the blended Fe-M \uta\ feature is very clear
\citep[e.g.,][]{sako01}. The present result raises the suspicion
that perhaps in other targets as well, this component of the
absorber is not part of the \agn\ outflow. Of course, this will
need to be checked in the future for each target separately.
However, if indeed this is the case, the ionization state
diagnosed for these absorbers based on the position of the \uta\
is underestimated, as the high velocities assumed for these
targets are being confused with higher charge states that are
actually moving at lower velocities.

Before concluding, we need to mention one disturbing caveat about
the present measurement. The rest frame wavelengths of the
resonant $n$~= 2 to 3 absorption lines, which we use to determine
the Fe-M outflow velocity, are taken from theoretical calculations
\citep{behar01}. Measuring these wavelengths in the laboratory is
rather challenging and indeed, they have not yet been confirmed.
As far as we know, Fe-M \uta\ absorption has been observed in the
laboratory once by \citet{cp00}. However, the spectral resolution
of that measurement was not sufficiently high to pin down the
positions of individual charge states.

The errors on the present velocity estimates stem directly from
the position uncertainties of the strongest lines in each \uta. At
16~\AA, an uncertainty in the rest frame wavelength of 30~m\AA\
translates into an uncertainty of 560~\kms\ in the outflow
velocity. Indeed, the scatter among wavelength calculations by
different state-of-the-art atomic codes, which is one way to get a
rough idea of their expected uncertainty, can reach and even
slightly exceed 30~m\AA. This, e.g., is the case for the
inner-shell K$\alpha$ absorption lines of oxygen
\citep{beharkahn02, schmidt04}. However, even if there are such
uncertainties associated with the Fe-M rest frame wavelengths that
we use, there is no particular reason for them to all be off in
the same direction as to indicate a uniformly erroneous velocity.
Quite the contrary, with the parametric potential method used by
\hullac\ \citep{bs01}, the atomic potential is calculated {\it ab
initio} and separately for each ion, so that errors can be
expected to be unconnected. Indeed, a comparison between EBIT
laboratory measurements and \hullac\ wavelengths for the strongest
line (the same 2p - 3d transition that dominates the Fe-M \uta s)
in each of the Fe-L ions reveals that $\lambda _{\mathrm {HULLAC}}
- \lambda _{\mathrm {EBIT}}$~= 0, $-$5, $+$6, $-$12, $+$8, $+$12,
$+$27, and $+$2 m\AA, respectively, for Fe$^{+16}$ -- Fe$^{+23}$
\citep{brown98, brown02}. Note that the laboratory wavelengths
themselves are associated with errors of a few m\AA.  In short,
the \hullac\ errors are generally much smaller than 30~m\AA\ and
are certainly not uniform.  Having said that, until the Fe-M \uta\
wavelengths are measured directly, this possibility can not be
rigorously ruled out. If it turns out that after all, the apparent
discrepancy between the outflow velocities of the Fe-M low charge
states and faster components of the outflow is due solely to the
inaccurate Fe-M atomic data we have used, then the \hetgs\ data
and Table~1 could serve as (the best currently available)
wavelength calibration for the Fe-M wavelengths.

\section{CONCLUSIONS}
\label{sec:concl}

We have measured the velocity of individual Fe-M ions in the \x\
outflow of \ngc\ for the first time. This measurement has never
been carried out before for \ngc\ nor for any other target. Based
on the theoretical rest frame wavelengths available to us, we find
that the Fe-M ions are not moving as fast as the more ionized Fe
species, such as Fe$^{+16}$, or as other highly charged species,
which comprise the high-velocity outflow of \ngc. In fact, the
Fe-M absorber may not be moving at all along our line of sight. We
speculate that this stationary gas may pertain to the ISM of \ngc,
to its narrow line region, or it may be gas evaporating off of the
skin of the molecular torus. In order to examine whether this is a
distinctive quality of \ngc\ or if there is a fundamental problem
with the atomic data, more high-quality spectra will need to be
obtained for similar targets, which show deep, resolved Fe-M \uta\
troughs.

\acknowledgments This research was supported by The Israel Science
Foundation (grant no. 28/03) and by grant no. 2002111 from the
United States Israel Binational Foundation. S.K. acknowledges the
financial support of the Zeff Fellowship.

\clearpage

\epsscale{.70}
\begin{figure}
%
\centerline{\includegraphics[width=13.0cm,angle=90]{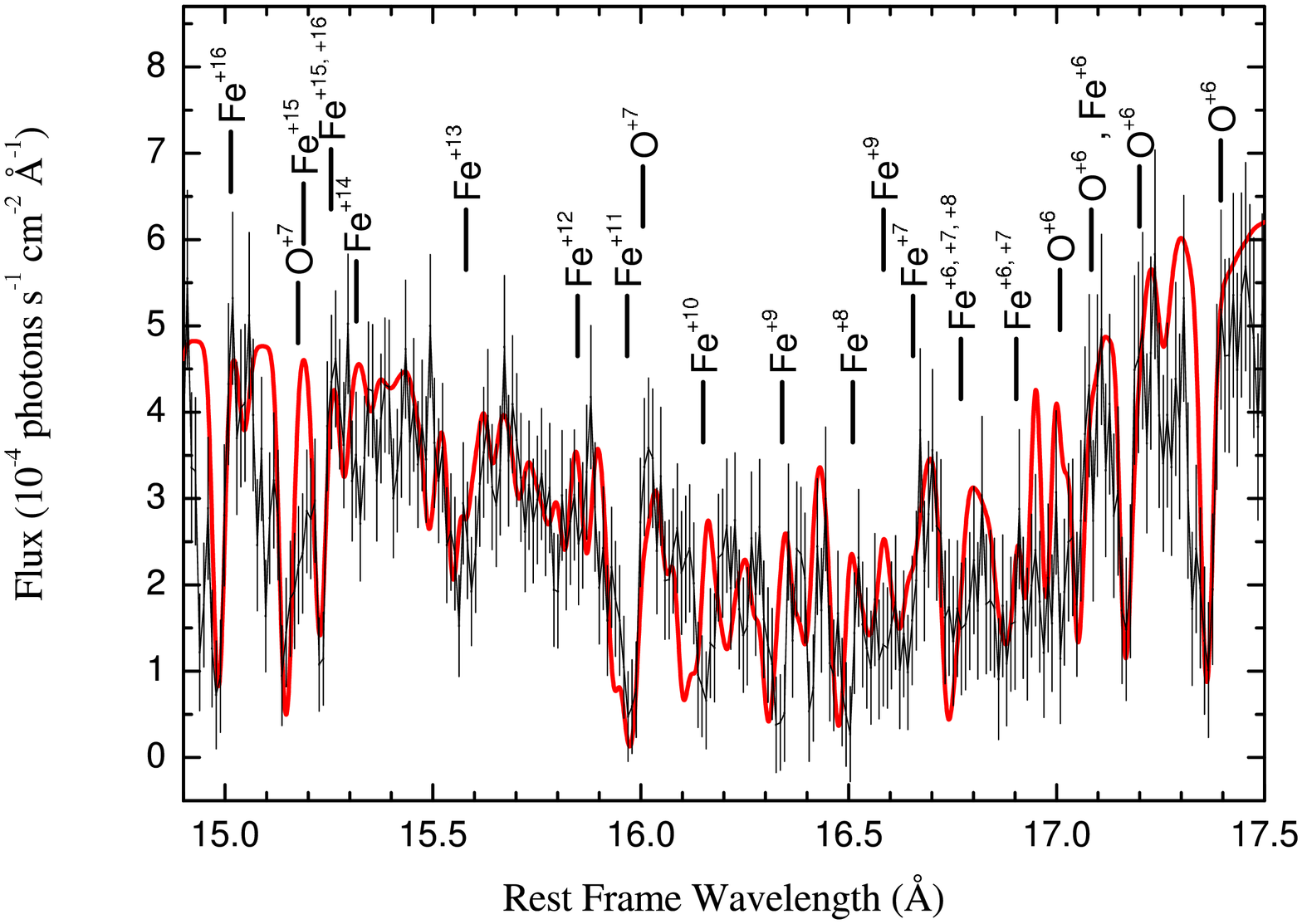}}
  \caption{\hetgs\ spectrum of \ngc\ in the Fe-M \uta\ region fitted with a
  uniform-velocity (--590 \kms) outflow model.
  The model fits some of the data very well, but not the Fe-M \uta\ features,
  most noticeably Fe$^{+8}$, Fe$^{+9}$, and Fe$^{+10}$ for which
  this velocity is clearly too high. }\label{f1}
\end{figure}

\clearpage

\begin{figure}
%
\centerline{\includegraphics[width=13.0cm,angle=90]{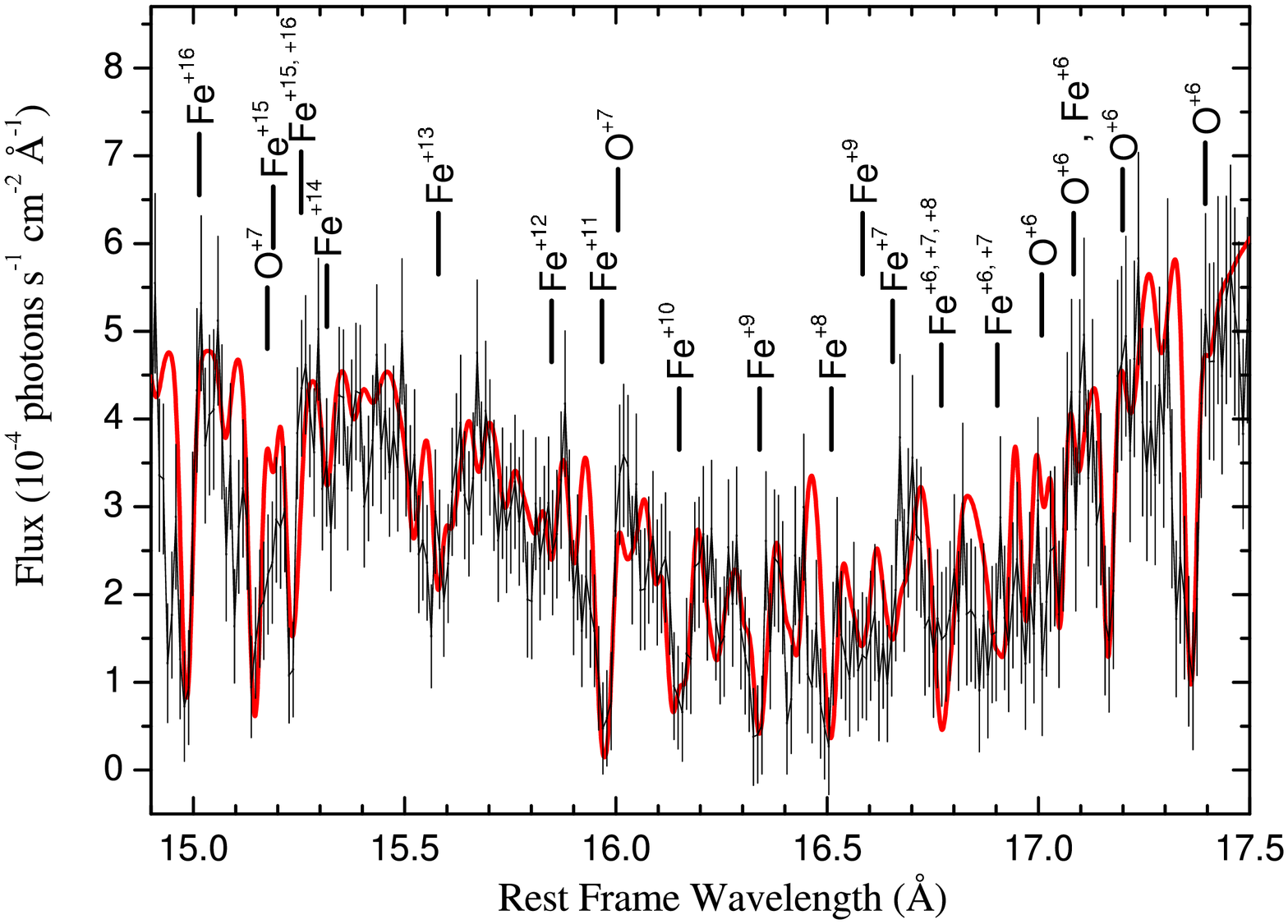}}
  \caption{\hetgs\ spectrum of \ngc\ in the Fe-M \uta\ region fitted with
  a model in which O$^{+6}$, O$^{+7}$, and Fe$^{+16}$ are outflowing at a
  velocity of $-590$ \kms, while the Fe-M ions are at rest with respect to the host galaxy.
  This model fits the observed Fe-M \uta\ features appreciably better than the uniform-velocity
  model shown in Fig.~1.} \label{f2}
\end{figure}

\begin{figure}
%
\centerline{\includegraphics[width=13.0cm,angle=0]{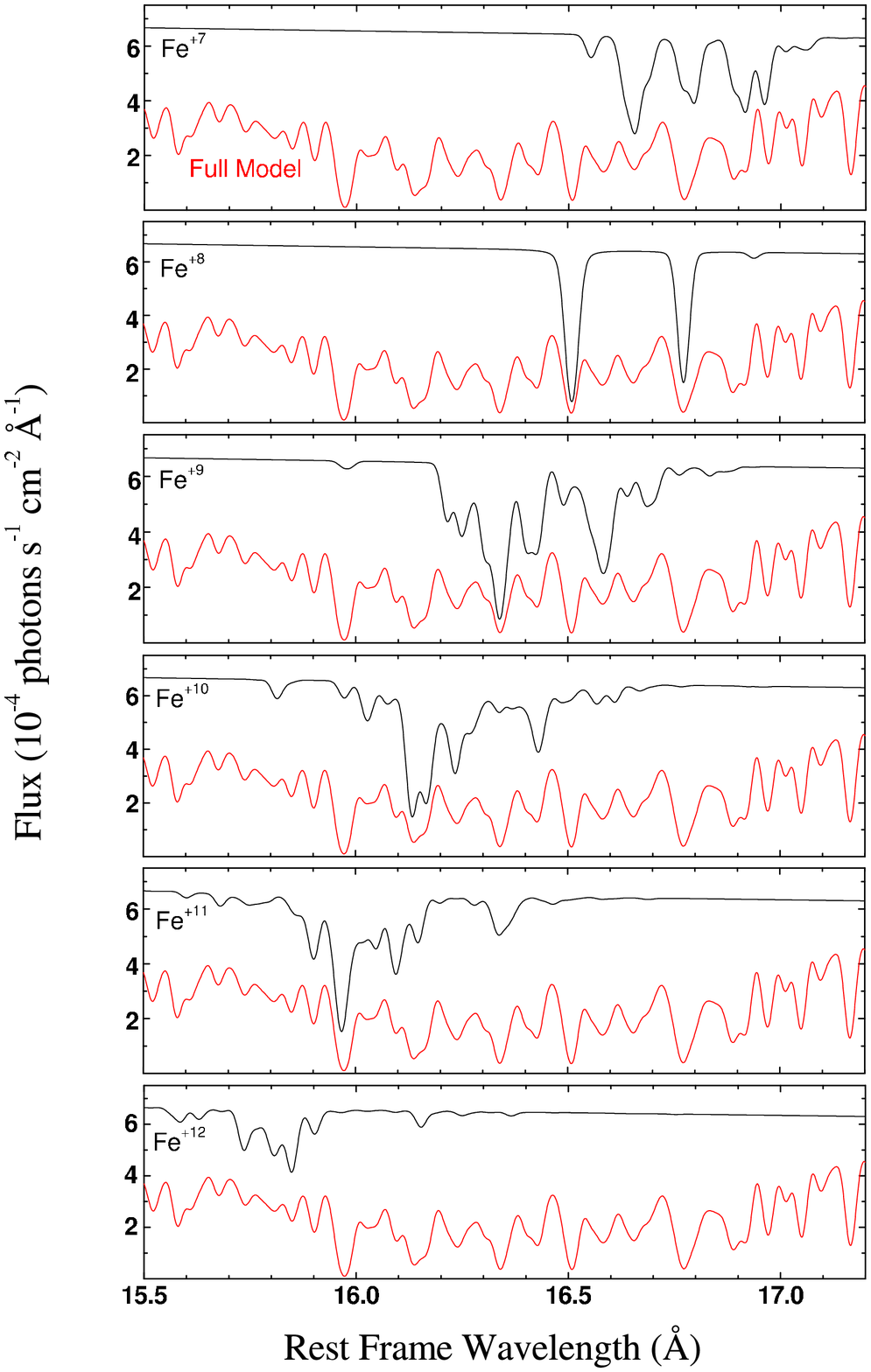}}
  \caption{Individual ionic contributions to the full best-fit model plotted in
  Fig.~2. For each ion, the strongest absorption feature can be easily discerned
  and its centroid position in the full model (red curve) is essentially unaffected by
  blends.}
   \label{f3}
\end{figure}

\end{document}